\documentclass[aps,prl,a4,twocolumn,preprintnumbers,nofootinbib]{revtex4}
\newcommand{\bi}{\bibitem}
\newcommand{\be}{\begin{eqnarray}}
\newcommand{\ee}{\end{eqnarray}}
\newcommand{\nn}{\nonumber}
\def\hbar#1{\slash\hspace{-2.5mm}#1}
\usepackage{epstopdf}
\usepackage{graphicx,graphics}% Include figure files
\usepackage{dcolumn}% Align table columns on decimal point
\usepackage{bm}% bold math
\begin{document}

\preprint{HIP-2008-10/TH}
\title{ Signatures for right-handed neutrinos at the Large Hadron Collider}
\author{Katri Huitu$^{1,}$}
%\email{katri.huitu@helsinki.fi}
\author{Shaaban Khalil$^{2,3,}$}%\email{skhalil@bue.edu.eg}
\author{Hiroshi Okada$^{2}$ }%\email{HOkada@bue.edu.eg}
\author{Santosh Kumar Rai$^{1,}$}%\email{santosh.rai@helsinki.fi}
%%%%%%%%%%%%%%%%%%%%%%%%%%%%%%%%%%%%%%
\affiliation{$^1$Department of Physics, University of Helsinki and Helsinki Institute of Physics, P.O. Box 64, FIN-00014 University of Helsinki, Finland.
}%
\affiliation{$^{2}$
Centre for Theoretical Physics, The British University in Egypt, El Sherouk City, Postal No, 11837, P.O. Box 43, Egypt.
}%
\affiliation{$^{3}$
Department of Mathematics, Ain Shams University, Faculty of Science, Cairo, 11566, Egypt.
}%
%%%%%%%%%%%%%%%%%%%%%%%%%%%%%%%
%\date{\today}
%%%%%%%%%%%%%%%%%%%%%%%%%%%%%%%
\begin{abstract}
We explore possible signatures for  right-handed neutrinos in TeV scale 
$B-L$ extension of the Standard Model (SM) at the Large Hadron Collider (LHC). 
The studied four lepton signal has a tiny SM background. We find 
the signal experimentally accessible at LHC for the considered parameter
regions.
\end{abstract}
%%%%%%%%%%%%%%%%%%%%%%%%%%%%%%%
%\pacs{.........}
\maketitle
%%%%%%%%%%%%%%%%%%%%%%%%%%%%%%%
The fact that neutrinos are massive indicates a firm evidence of
new physics beyond the Standard Model (SM). The most attractive
mechanism that can naturally account for the small neutrino masses
is the seesaw mechanism. In this case, three heavy singlet
(right-handed) neutrinos $\nu_{R_i}$ are invoked. Recently, a low
scale $B-L$ symmetry breaking has been considered, based on the
gauge group $SU(3)_C\otimes SU(2)_L\otimes U(1)_Y\otimes
U(1)_{B-L}$ \cite{Khalil:2006yi}. This model provides a natural
explanation for the presence of three right-handed neutrinos
and can account for the current experimental
results of the light neutrino masses and their mixings
\cite{Abbas:2007ag}. 

In $B-L$ extension of the SM, the right-handed neutrinos acquire
the following masses after the symmetry breaking: $M_{\nu_{R_i}}=
\frac{1}{\sqrt{2}} \lambda_{\nu_{R_i}} v^{\prime}$, where $v^{\prime}$
is the scale of $B-L$ breaking. Similar to the electroweak symmetry, 
the scale of $B-L$
can be linked to the supersymmetry breaking scale at the
observed sector, with the $B-L$ symmetry radiatively
broken at TeV scale \cite{masiero}. Thus, the right-handed
neutrino mass can be of order ${\cal O}(100)$ GeV, depending on
the value of the Yukawa couplings $\lambda_{\nu_{R}}$ which
augurs well for its direct search at the Large Hadron Collider (LHC). 
In addition, one extra neutral
gauge boson ($Z'$) corresponding to $B-L$ gauge symmetry is
predicted in this type of models. This gauge boson couples to both
the SM fermions and right-handed neutrinos through the
non-vanishing $B-L$ quantum numbers and gives the dominant 
contribution to the production of the
right-handed neutrino at LHC. It is worth noting that in the SM
extended with right-handed neutrinos, this production at LHC \cite{rnu-lhc}, 
is mainly through the exchange of $W$ boson, and is thus suppressed by the
small mixing between light and heavy neutrinos.

The aim of this letter is to analyze the LHC discovery potential for
the lightest right-handed neutrino in TeV scale $B-L$ extension of the
SM. We provide a detailed phenomenological analysis for such a
neutrino and show that the right-handed neutrinos are
accessible via a clean signal at LHC. Our results indicate that
observation of $\nu_R$ signals at LHC would significantly
distinguish between TeV scale $B-L$ extension of the SM and other
scenarios for SM extended with right-handed neutrinos.

In the minimal version of the $B-L-$type extension of the SM, the
interactions between right-handed neutrino and matter fields
are described by the Lagrangian %
\be%
{\cal L}_{\nu_R} &=& i{\bar \nu_R}D_{\mu}\gamma^{\mu}\nu_R -
(\lambda_{\nu}{\bar l} {\tilde \phi}\nu_R +
\frac12\lambda_{\nu_R}{\bar\nu}^c_R\chi\nu_R+h.c.) \nn
\\
&-& V(\phi,\chi) \label{lagrangian} \ee
where the covariant derivative $D_{\mu}$ is defined as $D_{\mu}
\nu_R = (\partial_{\mu} - i g'' Y_{B-L}Z'_{\mu}) \nu_R$, where
$g''$ is the $U(1)_{B-L}$ gauge coupling constant and $Y_{B-L}$ is
the corresponding $B-L$ charge. $\lambda_{\nu}$ and
$\lambda_{\nu_R}$ refer to the $3 \times 3$ Yukawa matrices. The
scalar potential $V(\phi,\chi)$ for the two scalars $\phi$ and $\chi$
is defined by \cite{Khalil:2006yi}%
\be%
V(\phi, \chi) &=& m^2_1\phi^{\dagger}\phi +
m^2_2\chi^{\dagger}\chi+ \lambda_1(\phi^{\dagger}\phi)^2 +
\lambda_2(\chi^{\dagger}\chi)^2\nn\\
&+& \lambda_3(\phi^{\dagger}\phi)(\chi^{\dagger}\chi),
\label{potential}%
\ee%
where $\lambda_3 > -2 \sqrt{\lambda_1 \lambda_2}$ and $\lambda_1$,
$\lambda_2 \geq 0$ so that the potential is bounded from below. The field
$\phi$ is the usual SM doublet while $\chi$ is a SM singlet complex scalar 
field, responsible for the spontaneous breaking of the $B-L$ symmetry.
After the breakdown of the $B-L$ and electroweak symmetry, 
mixing between $\phi$ and $\chi$ and also between $\nu_L$ and $\nu_R$ are
generated. These mixings initiate new interactions between the
right-handed neutrinos and the SM particles.

The mixing between the neutral scalar components of Higgs multiplets,
$\phi^0$ and $\chi^0$, leads to the following mass
eigenstates, which we define as $H$ (SM-like Higgs boson) and $H'$ (heavy
Higgs boson):
\be%
\left(\begin{array}{c}
 H\\
 H' \\
\end{array}\right)
&=&
\left(\begin{array}{cc}
\cos\alpha & -\sin\alpha \\
 \sin\alpha  & \cos\alpha \\
\end{array}\right)
\left(\begin{array}{c}
 \phi^0\\
 \chi^0 \\
\end{array}\right),
\ee %
where $\alpha$ is the Higgs mixing angle which is given by
\cite{s.khalil2}%
\be%
\tan2\alpha=\frac{|\lambda_3|vv'}{\lambda_1 v^2-\lambda_2 v'^2},%
\ee%
while $v$ and $v'$ are the vacuum expectation values (VEV) given to
$\phi$ and $\chi$ respectively.
On the other hand, the mixing between $\nu_L$ and $\nu_{R}$ can
be represented by the following $6\times 6$ mass matrix: %
\be %
M(\nu_L,\nu_R) &=& \left(\begin{array}{cc}
 {\bf 0} & m_D^T \\
 m_D & M_R
 \label{Mnu}
\end{array}\right),
\ee%
where $m_D\sim\lambda_{\nu}v$ is Dirac mass term  and
$M_R\sim\lambda_{\nu_R}v'$ is Majorana mass term for neutrinos. Therefore, the
mass eigenstates $\nu_l$ (light neutrinos) and $\nu_h$
(heavy neutrinos) are given by%
\be%
\left(\begin{array}{c}
\nu_l\\
\nu_h \\
\end{array}\right)= \left(\begin{array}{cc}
 U & - U V \\
 V^T & {\bf 1} \\
\end{array}\right)
%%%%%%%%%%%%%%%
\left(\begin{array}{c}
 \nu^c_L\\
 \nu_R \\
\end{array}\right)
\label{mass-state}. \ee
%%%%%%%%%%%%%%%%%
Here the matrix $U$ refers to Maki-Nakagawa-Sakata mixing matrix
in light neutrino sector and the matrix $V$ is given by $V= m_D^T
M_R^{-1}$. It is worth mentioning that the mass eigenstates in
Eq.(\ref{mass-state}) are obtained by applying two successive
rotations. The first one transforms the mass matrix
$M(\nu_L,\nu_R)$ in Eq.(\ref{Mnu}) to $diag\{m_{\nu}^{eff},
M_R\}$, where $m_{\nu}^{eff}= m_D^T M_R^{-1} m_D$ which is
diagonalized by the $U$ matrix.  We adopt the Dirac neutrino mass
matrix $m_{D}$ as found in Ref.\cite{Abbas:2007ag} from the
extended mass relations among the quark and lepton masses. In this
example, $m_D$ is non-hierarchical matrix with entries of order
$10^{-4}$ GeV. Since we assume that $M_{\nu_{R_1}} \ll M_{\nu_{R_2}} <
M_{\nu_{R_3}}$, the resultant mixing matrix $V$ is characterized
by the following feature: $V_{11} \gg V_{1i}$ for $i=2,3$. Also
the typical value of $V_{11}$ is of order $10^{-6}$. Although this
mixing is rather small, it generates new coupling between the heavy
neutrino, the weak gauge bosons $W$ and $Z$, and the associated
leptons. This new coupling plays
an important role in the decays of the lightest heavy neutrino
($\nu_{h_1}\equiv N_1$).

%%%%%%%%%%%%%%%%%%%%%%%%%%%%%% Dominant Decay Mode Figure %%%%%%%%%%%%%
\begin{figure}[t]
%\vskip -2.0cm
\includegraphics[height=5.5cm,width=5.5cm]{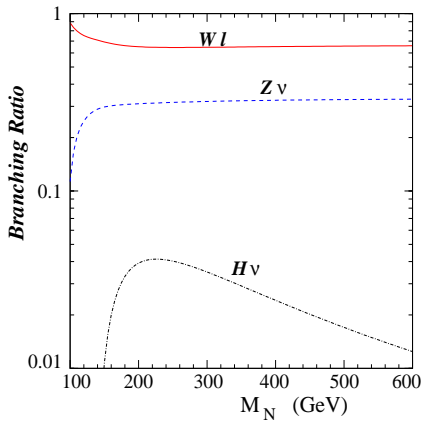}%Here is how to import EPS art
\caption{\label{dominant-decay} The branching ratio for the various 
decay modes of the right-handed neutrinos.}
\end{figure}
%%%%%%%%%%%%%%%%%%%%%%%%%%%%%% Dominant Decay Mode Figure %%%%%%%%%%%%%
Now, we can express the relevant interactions that lead to
dominant contributions to the production and decay of the lightest
heavy neutrino $N_1$ at LHC:%
\be 
\label{eq:lint}
{\cal L}_I \sim &-& g''Z'_{\mu} [\overline{N}_1\gamma^{\mu}N_1 +
(UV)_{i1}\overline{(\nu_l)}_i\gamma^{\mu} N_1 + h.c.] \nn \\
%%%
&+& \frac{g_2}{2c_W}Z_\mu(UV)_{i1}\overline{(\nu_l)}_i\gamma^{\mu} N_1 \nn \\
&+& \frac{g_2}{\sqrt2} V_{i1}W^-_{\mu}l^{+}_i\gamma^{\mu} N_1 +h.c.,%
\ee%
where the family index $i=1,2,3$. From this interaction 
Lagrangian, one finds that the dominant production mode for the
heavy neutrino $N_1$ is through the exchange of $Z'$ gauge boson
and the main decay channel is through $W$ gauge boson
as shown in Fig. \ref{dominant-decay}.

Some comments are in order: $(i)$ In SM extended with right-handed neutrinos,
there is no extra gauge boson and hence the production of
right-handed neutrinos may be obtained via the exchange of $Z$ or
$W$ only with a suppression factor due to the mixing between light
and heavy neutrinos. $(ii)$ The decay modes for the $N_1$ depend
on the Yukawa strength $\lambda_\nu$ and the mixing parameter $V_{11}$.
Since both are of the order of $10^{-6}$ as pointed out earlier, we find that
the most dominant decay modes are $W^+~e^-$ and $Z~\nu_e$, with a small 
fraction into $H~\nu_e$. In our
analysis, we find that 
%for $V_{11}  \simeq \lambda_\nu = 10^{-6}$ 
the BR$(N_1\to W^+e^-)$ is always dominant, ranging between 0.65-0.89 while
BR$(N_1\to Z~\nu_e)$ is 0.11-0.33 for 
$V_{11} \simeq 2 \times (\lambda_\nu = 10^{-6})$, 
for right-handed neutrino masses $M_N > 100$ GeV.

\begin{figure}[htb]
\includegraphics[width=6cm,height=6cm]{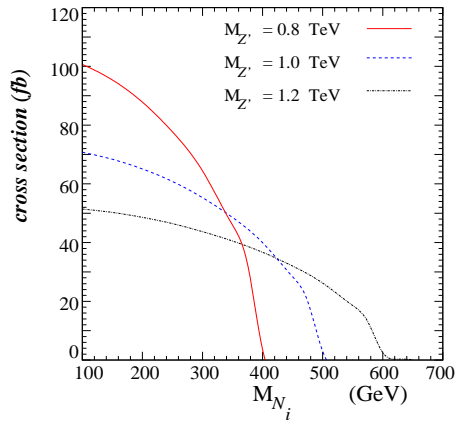}%Here is how to import EPS art
\caption{\label{fig:epsart1} Illustrating the pair production cross
section for the right-handed neutrinos at LHC. }
\end{figure}

As mentioned, the dominant production mode for the right-handed 
neutrinos at the LHC would be through the Drell-Yan
mechanism, with $Z'$ in the $s$-channel. The new gauge
quantum number associated with the $B-L$ symmetry couples 
the right-handed neutrinos directly to the gauge boson $Z'$, 
as seen from the Lagrangian in Eq.(\ref{eq:lint}). Thus
the rate for the pair production of the heavy neutrinos would
crucially depend on the mass of the $Z'$  and the strength of
the $B-L$ coupling $g''$. In Fig. \ref{fig:epsart1} we plot the
pair production cross section for a pair of right-handed neutrinos
at the LHC, as a function of the right-handed neutrino mass ($M_N$) for
three different choices of the $Z'$ mass ($M_{Z'}$).
%%%%%%%%%%%%%%%% Cross Section for nR %%%%%%%%%%%%%%%%%%%
%%%%%%%%%%%%%%%%%%%%%%%%%%%%%%%%%%%%%%%%%%%%%%%%%%%%%%%%%
The $B-L$ coupling and the $Z'$ mass is chosen in such a
way that it always respects the LEP bound \cite{lep-ii}. Furthermore, the 
recent results by CDF II \cite{Abulencia:2006iv} are consistent with the 
LEP II constraints in case of $B-L$ extension of the SM, with a typical
lower bound ${M_{Z'}}/{g''} > 6$ TeV. 
We choose benchmark points of the model for our analysis, as given below :
%%%%%%%%%%%%%%%%%%%%%%%%%%%%%%%%%%%%%%%
\begin{eqnarray}
&& \lambda_1 = 0.15,~ \lambda_2 = 0.02,~ \lambda_3 = -0.001, \nonumber \\
&& v = 246~ {\rm GeV},~ v^\prime = 3~ {\rm TeV}, 
~\lambda_\nu = 10^{-6},~ V_{11} = 2\times 10^{-6} \nonumber, \\
&& (g'',M_{Z^\prime}) = (0.133,800),~(0.167,1000),~ (0.2,1200) .\nonumber 
\end{eqnarray}
The production cross section is enhanced due to the resonant contribution
from the $Z'$ exchange in the $s$-channel, but falls rapidly with increasing
right-handed neutrino mass. We now focus on the event rates for the most
promising signal coming from the pair production of the right-handed neutrinos
in this model.
We choose two points from Fig. \ref{fig:epsart1} to highlight the signal
for the right-handed neutrinos at LHC, {\it viz.}
($g'' = 0.133,~M_{Z^\prime} =  800$ GeV, $M_N=200$ GeV) and
($g'' = 0.2,~M_{Z^\prime} = 1200$ GeV, $M_N=400$ GeV).
The right-handed neutrinos dominant decays are to a $W$-boson and a charged 
lepton and to a left-handed neutrino and the $Z$ boson,
through the mixing parameter $V_{ij}$. 
%For very small mixing 
%parameter $V_{11} << \lambda_\nu$, the dominant decay for the right-handed 
%neutrino would be to a light Higgs and neutrino. This in itself would
%be a very interesting signal, where one has two Higgs bosons with large 
%missing $E_T$, provided the rates are high. We, however, choose to discuss the
%signal coming from the other decay modes ($Z\nu,Wl$), as it is much cleaner 
These decays are very clean with four hard leptons in the final states and 
large missing energy due to the associated neutrinos.
%%%%%%%%%%%%%%%%%%%%%%%%%%%%%%%%%%%%%%%%%%%%%%%%%%%%%%%%%
%\begin{widetext}
%\begin{center}
\begin{figure}[ht]
\includegraphics[height=5.5cm,width=8.5cm]{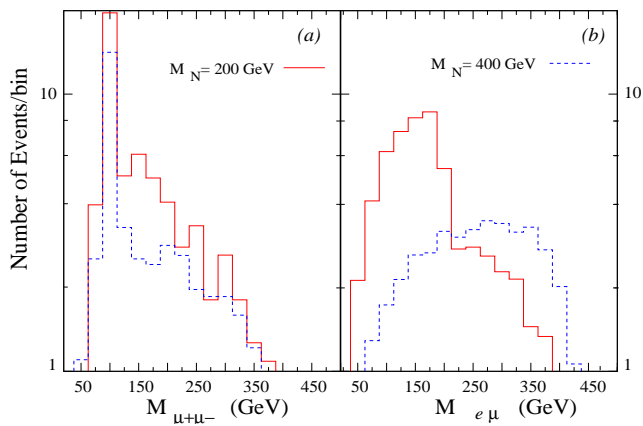}
\caption{\label{fig:epsart2}
Illustrating the bin-wise distribution in the invariant mass of the
charged lepton pairs in the final state for two different $N_1$ masses,
$M_N=200$ GeV and $M_N=400$ GeV.}
\end{figure}
%\end{center}
%\end{widetext}
%%%%%%%%%%%%%%%%%%%%%%%%%%%%%%%%%%%%%%%%%%%%%%%%%%%%%%%%%
The SM background for such a final state is negligible at the LHC.
The dominant processes in the SM come through $4W$ productions with
$\sigma(4W)\sim 6~fb$ \cite{barger:1989}, and the dominant contributions
coming from the three gauge boson $WWZ$ productions at LHC, with 
$\sigma(WWZ)\sim 200~fb$ (including QCD corrections)\cite{Hankele:2007sb}. 
However, because of the smallness of the pure leptonic branching ratios, the
cross section of the four lepton final states fall to $\mathcal{O}(10^{-4})~fb$ 
and $\mathcal{O}(10^{-2})~fb$ for the $4W$ and $WWZ$ modes respectively. This
is further rendered negligible once we demand the minimum acceptance cuts 
on the kinematic variables for our signal.
 
Depending on the production and decay mechanism, we can have the following
final states as our signal:
%\begin{itemize}
%\item 
$$l^+_i l^-_j l^+_k l^-_m+{\hbar E}_T+X$$
%\item $l^+_il^-_i+l^+_jl^-_j+{\hbar E}_T+X ~~~ (i\neq j)$,
%\end{itemize}
where $i,j,k,m$ run over the three different lepton flavors. 
We put the following
kinematic cuts when selecting the final states for our analysis:
%\begin{itemize}
(a) For the charged leptons: $p_T^l > 20$ GeV and  $|\eta| < 2.5$.
(b) A minimum missing transverse energy(momentum) cut ${\hbar E}_T > 50$ GeV.
(c) For resolving the different leptons in the detector,
$\Delta R_{l_i l_j}\ge0.2$ and a minimum cut on the invariant mass 
$M_{l_i^+ l_i^-} > 10$ GeV.
%\end{itemize}

To calculate and generate the events, we include the relevant couplings of the
model in CalcHEP 2.4.5 \cite{calchep} and pass the generated events through the
CalcHEP+Pythia interface. We have turned on the initial and final state 
radiation effects using the
Pythia \cite{pythia} switches. We use the leading order CTEQ6L \cite{cteq}
parton distribution functions (PDF) for the protons colliding at LHC. 
%%%%%%%%%%%%%%%%%%%%%%%%%%%%%%%%%%%%%%%%%%%%%%%%%
\begin{figure}[t]
\includegraphics[height=5.5cm,width=8.5cm]{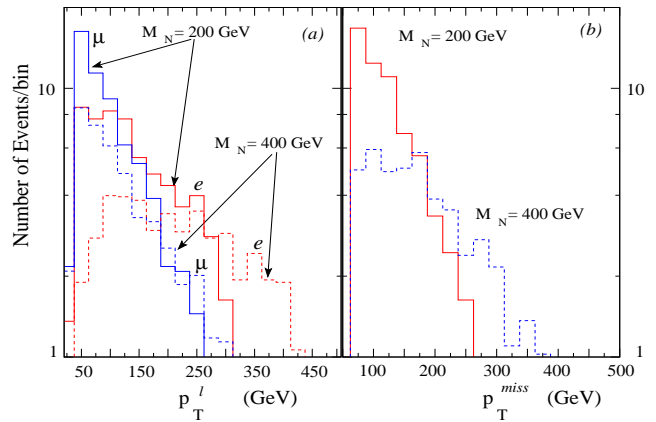}%Here is how to import EPS art
\caption{\label{fig:epsart3}
Illustrating (a) the transverse momentum distribution for the different
charged leptons and (b) the missing transverse momentum, for
the 4$l$+${\hbar E}_T$ signal at the LHC for two different $N_1$ masses,
$M_N=200$ GeV and $M_N=400$ GeV.}
\end{figure}
%%%%%%%%%%%%%%%%%%%%%%%%%%%%%%%%%%%%%%%%%%%%%%%%%
In Fig. \ref{fig:epsart2} and \ref{fig:epsart3}, we plot the various 
kinematic distributions for the signal arising through
\begin{eqnarray*}
pp\rightarrow N_1 \bar{N}_1\rightarrow e^+e^-\mu^+\mu^-{\hbar E}_T,
\end{eqnarray*}
satisfying the above selection cuts. We choose an integrated luminosity 
of $\int \mathcal{L}dt$ = 300 $fb^{-1}$.
As the dominant decay of $N_1$ is to $W$-boson and electron,
the $\mu's$ would most often come from the $Z$ decay, when one right-handed 
neutrino decays via the $W$ mode and the other through the $Z$ mode. Thus one 
expects a clear peak at $M_Z$, mass of the $Z$-boson in the 
$M_{\mu^+\mu^-}$ distribution as seen in
Fig. \ref{fig:epsart2}(a). The invariant mass of the electrons is more wide as 
compared to the invariant mass of the muons. With the SM background completely
reducible this gives a clear information on which neutrino
flavor is produced in the $pp$ collisions. Another
interesting feature is seen in Fig. \ref{fig:epsart2}(b), where we plot the
invariant mass of the different flavor leptons. A distinct kinematic edge is
seen at $M_{e\mu}\simeq M_{N_1}$ which is mainly because of the large 
contributions coming from the dominant decay through the $W$ boson, following 
the decay chain $N_1 \to e W^* \to e \mu \nu_\mu$. A more efficient way of
identifying the edge would come if one selects an invariant mass window for
$(M_Z-10 ~{\rm GeV}~< M_{\mu^+\mu^-} < M_Z+10~{\rm GeV})$ and looks at the 
invariant 
distribution of $M_{e^+e^-}$. This would correspond to the scenario where the
muon pairs always come from $Z$ whereas the electron pairs come from the 
cascade of $N_1$. The $M_{e^+e^-}$ distribution would then show a clear sharp
edge at $M_{N_1}$ and thus give a very precise determination of the mass of the
right-handed neutrino albeit we have a smaller event rate.

%%%%%%%%%%%%%%%%%%%%%%%%%%%%%%%%%%%%%%%%%%%%%%%%%
Fig. \ref{fig:epsart3}(a) shows that the electrons coming from the 
primary decay of the right neutrino are much harder than the muons coming 
from the lighter weak gauge bosons. The missing $p_T$ distribution 
in Fig.\ref{fig:epsart3}(b) represents the light neutrinos in 
the final state. With an integrated luminosity of 
300 fb$^{-1}$ the expected number of events for the $4 \ell ~{\hbar E}_T$ final 
states is 71 when $M_{N_1}=200$ GeV and 46 when $M_{N_1}=400$ GeV. The $4\ell$ 
can be either $4e,3e~1\mu,2e~2\mu$ or $1e~3\mu$ depending on the decays. 
However, even with an integrated luminosity of 30 fb$^{-1}$, we still 
expect 5-7 events for the above final state. This is quite an encouraging 
result where we have negligible SM background. 

The situation gets more complicated if two flavors of the right-handed
neutrinos are assumed to be degenerate in mass. Then one has the same final
states for both $N_1$ and $N_2$ pair production with similar event
rates. This would result in loss of the clear correlation that existed between
the different charged
lepton flavors as shown in the various kinematic distributions, rendering it 
difficult to exploit the advantages which were
perceivable in the invariant mass distributions. However, one advantage would
be the doubling of the total number of events in the final state. 
%One can also
%simultaneously look for the other final states, like for instance
%($4e~{\hbar E}_T+X$) and ($4\mu~{\hbar E}_T+X$) or ($3e~1\mu~~{\hbar E}_T+X$) 
%and ($3\mu~1e~~{\hbar E}_T+X$). 
Other promising signatures arise
from the pair production of right-handed neutrinos in this model at the LHC,
if the $W$ and $Z$ bosons were allowed to decay 
hadronically \cite{delAguila:2007}. 
This would give ($3\ell+2j+{\hbar E}_T$) or ($2\ell+4j$) in the final 
state. Being a hadron machine, any final state with jets will have a large QCD 
background. However, with a selection window of 20 GeV around the weak gauge 
boson masses for the 2-jet invariant mass, one can reduce a large part of 
the SM background.

Finally, let us note that a $4\ell+{\hbar E}_T$ final state is
possible also in other beyond the standard model scenarios,
like for e.g. in supersymmetric theories \cite{Baer:2000pe}. The
signal in supersymmetric theories can come from pair production of 
heavy neutralinos, heavy Higgs bosons \cite{Bisset:2007mi} which can give
comparable and even larger event rates when compared to our case.  
%one should not ignore the possibility of other new physics 
%scenario giving a $4\ell+{\hbar E}_T$ final state, like for e.g. in 
%supersymmetric theories \cite{Baer:2000pe}. 
However, the invariant mass 
distribution for the charged lepton pairs can very effectively distinguish
our scenario. The distinct kinematic edge seen in the $e-\mu$ distribution
and the $Z$-peak in the $\mu^+\mu^-$ distributions shown in 
 Fig. \ref{fig:epsart2}(b) and (a) respectively will not appear in the
supersymmetric case, where the kinematic edge will be seen in the invariant 
mass distribution of the oppositely charged leptons of same flavor
\cite{Baer:2000pe,Cheung:1997tg,Bisset:2007mi}.

%%%%%%%%%%%%%% Section V %%%%%%%%%%%%%%%%%%%%%%%%%%%%%
%\section{\label{sec:level5}Discussions and Conclusions}
In this letter we have considered the TeV scale $B-L$ extension of
the SM. We provided a comprehensive analysis for the phenomenology
of the (heavy) right-handed neutrinos with $U(1)_{B-L}$ charge.
We find that the production rate of the right-handed neutrinos is quite large 
over a significant range of parameter space. Searching for the right-handed 
neutrinos is accessible via a very clean signal at LHC, with negligibly small
SM background. We also find a distinct correlations among the final state
leptons coming from the decay of the lightest right-handed neutrinos.

\vspace*{0.2cm}
\begin{acknowledgments}
We thank J. A. Aguilar-Saavedra for helpful comments. 
SKR would like to thank A. Belyaev for help in clearing some doubts on the
Calchep-Pythia interface. 
KH and SKR gratefully acknowledge support from the Academy of
Finland (Project No. 115032).
\end{acknowledgments}

\vskip -0.5cm
%\newpage %Just because of unusual number of tables stacked at end
\bibliography{Signatures for right handed neutrinos at the Large Hadron Collider}% Produces the bibliography via BibTeX.

\end{document}